\documentclass{article}
\usepackage[utf8]{inputenc}
\usepackage[english]{babel}
\usepackage{graphicx}
\usepackage{bm}
\usepackage[varg]{txfonts} 
\usepackage[margin=0.7in]{geometry}
\begin{document}

\title{A general vector field coupled to a strongly compressible turbulent flow}

\author{N. V. Antonov$^{1,2}$ and M. M. Tumakova$^3$}

\date{$^1$ Department of Physics, Saint Petersburg State University, \\
7/9 Universitetskaya Naberezhnaya, Saint Petersburg 199034, Russia; \\
$^2$ N.N. Bogoliubov Laboratory of Theoretical Physics, Joint Institute for Nuclear Research, Dubna 141980, Moscow Region, Russia; \\
$^3$  Landau Institute for Theoretical Physics, RAS,
 Chernogolovka 142432, Moscow Region, Russia }

\maketitle

\begin{abstract}
We consider the model of a transverse vector (e.g. magnetic) field  with the most general 
form of the nonlinearity, known as the ${\cal A}$ model, 
passively advected by a strongly compressible turbulent flow, governed by the
randomly stirred Navier-Stokes equation. The full stochastic problem is equivalent to a certain renormalizable field theoretic model with an infrared attractive fixed point.
Thus, the scaling behaviour for the large-scale, long-distance behaviour is established.
However, the question whether the parameter ${\cal A}$ tends to a certain fixed-point value of the renormalization group equations or remains arbitrary, cannot be answered within the one-loop approximation of our study.\\

\end{abstract}

\section{Introduction}

In a few last decades of the past Millennium, 
intermittent interest had been attracted to the problem of intermittency and anomalous scaling in fluid turbulence; see e.g. Refs. \cite{Frisch,FGV}  and the literature cited therein. 

The term ``anomalous scaling'' reminds of the critical scaling in models of equilibrium phase transitions. In those, the field theoretic methods were successfully employed to establish the existence of asymptotic scaling regimes and to construct regular perturbative schemes (the famous $\varepsilon$ expansion and its relatives) for the corresponding exponents, scaling functions, etc; see e.g. \cite{Book} and references therein. 

In turbulence, the phenomenon manifests itself in singular (arguably power-like) behavior of various statistical quantities as functions of the integral turbulence scales, with infinite sets of independent anomalous exponents \cite{Frisch}. 

Within the framework of numerous semi-heuristic approaches, the anomalous exponents were related to statistical properties of the local dissipation rate, the fractal dimension of structures formed by the small-scale turbulent eddies, the characteristics of nontrivial structures 
(vortex filaments), and so on; see Refs. \cite{Frisch,Multy,Multy1,Multy2} 
for a review and further references. 
In spite of this undeniable achievement\footnote{Giorgio Parisi was awarded 
the Nobel Prize in Physics 2021 ``for the discovery of the interplay of disorder and fluctuations in physical systems'' \cite{Nobel}.},
the problem still remains open to further investigation.

The common feature of such models is that they are only intuitively related to underlying dynamic equations, involve arbitrary adjusting parameters and, therefore, cannot be considered to be the firm basis for of a systematic quantitative expansion 
in a certain small parameter. 
No regular calculation scheme, based on an underlying dynamical model and reliable perturbation expansion (hopefully similar to the famous $\varepsilon$ expansion for critical exponents) had yet been constructed for the anomalous exponents of the turbulent velocity field.
Thus, some doubt still remains about the universality of anomalous exponents and the very existence of deviations from the classical K41 theory.

In this direction of study, 
an extremely inspiring  progress was achieved for the simplified Kraichnan's rapid-change model of a passively advected scalar field
\cite{Kr68}, where the advecting velocity field was taken Gaussian, not correlated in time, and having a power-like correlation function 
of the form $\sim \delta (t-t')/k^{d+\xi}$, where $d$ is the dimension of space, $k$ being the wave number and $\xi$ an arbitrary exponent. There, for the first time, the existence of anomalous scaling was firmly established on the basis of a dynamic model; the corresponding anomalous exponents were calculated in controlled approximations \cite{GK,GK1,GK2} and, eventually, within a systematic perturbation expansion in a formal small parameter $\xi$ \cite{AAV,AAV2} 
up to the order $\xi^3$; see \cite{cube,cube2} and the references.

In the original Kraichnan's ``rapid-change model'', the velocity ensemble was taken Gaussian, not correlated in time, isotropic, and the fluid was assumed to be incompressible. More realistic models should take into account finite correlation time and non-Gaussianity of the velocity ensemble, anisotropy, compressibility of the fluid, etc; see the discussion in \cite{ANT500}.

The most efficient way to study anomalous scaling is provided by the field theoretic renormalization group (RG) combined with the operator product expansion (OPE); see \cite{Book,UFN,RedBook} for the detailed exposition of these techniques and the references. In the RG+OPE scenario for the anomalous scaling in turbulence, proposed in \cite{JETP}, the singular dependence on the integral scales emerges as a consequence of the existence in the corresponding models of composite fields with negative dimensions (termed in \cite{JETP} ``dangerous operators'').

In the RG+SDE treatment of the 
 Kraichnan's model, the anomalous exponents are identified with the scaling dimensions (``critical dimensions'' in the terminology of the critical state theory) of certain {\it individual} Galilean-invariant composite operators \cite{AAV}. This allows one to give a self-consistent derivation of the anomalous scaling, to construct a systematic perturbation expansion for the anomalous exponents in $\xi$, and to calculate the exponents in the second \cite{AAV,AAV2} and in the third \cite{cube,cube2} orders. The RG approach can be generalized to the case of finite correlation time \cite{ANT1} and to the non-Gaussian advecting velocity field, governed by the stochastic Navier-Stokes (NS) equation \cite{ANT3}. A general overview of the RG approach to Kraichnan's model and its descendants and more references can be found in \cite{ANT500}.

A next important step toward the real NS turbulence is to consider the turbulent advection of passive {\it vector} fields. The latter can have different physical meaning: magnetic field in the Kazantsev-Kraichnan model of magnetohydrodynamical turbulence in the kinematic approximation; perturbation in the linearized NS equation with prescribed statistics of the background field; density of an impurity with internal degrees of freedom, etc. Despite the obvious practical significance of these physical situations, the passive vector problem is especially interesting because of the insight it offers into the inertial-range behavior of the NS turbulence. 

Owing to the coupling between different components of the vector field (both by the dynamical equation and the incompressibility condition) and to the presence of a new stretching term in the dynamical equation, that couples the advected quantity to the gradient of the advecting velocity, the behavior of the passive vector field appears much richer than that of the scalar field: ``...there is considerably more life in the large-scale transport of vector quantities,'' (p. 232 of Ref. \cite{Frisch}). 
Indeed, passive magnetic fields reveal anomalous scaling already on the level of the pair correlation function \cite{Verg,Rog}. 

The general vector ``${\cal A}$ model'' introduced in \cite{1}, and further studied in \cite{2}--\cite{4},
includes as special cases the kinematic magnetic model
(${\cal A}=1$), linearized NS equation (${\cal A}=-1$) and the special model without the stretching term (${\cal A}=0$). Thus, in particular, it allows one to control the nonlocal pressure contribution ($\sim ({\cal A}-1)$)
and to quantitatively study its effects on the anomalous scaling. The generalized model also naturally arises within the multiscale technique, as a result of the vertex renormalization \cite{Frisch}.

Other important issues are the mixing of composite operators, responsible for the anomalous scaling, and the effects of pressure on the inertial-range behavior, especially in anisotropic sectors. In the scalar case, the anomalous exponents for all structure functions are given by a single expression which includes $n$, the order of the function, as a parameter \cite{AAV}. This remains true for the general vector model with the stretching term (${\cal A}\ne0$), 
including the magnetic case \cite{LM}-\cite{5} with ${\cal A}=1$.

In the special vector model without the stretching term (${\cal A}=0$), introduced and studied  in \cite{Runov}-\cite{AGK22}, the number and the form of the operators entering into the relevant family depend essentially on $n$, and different structure functions should be studied separately. As a result, no general expression valid for all $n$ exists in the model, and the anomalous exponents are related by finite families of composite operators rather than by individual operators. In this respect, such models are closer to the nonlinear NS equation, where the inertial-range behavior of structure functions is believed to be related with the Galilean-invariant operators, which form infinite families that mix heavily in renormalization. 

It was argued that similar mechanism is responsible for the origin of anomalous scaling in the real fluid turbulence, see, e.g. \cite{UFN,RedBook,JETP}.

Numerous studies were devoted to the joint effects of compressibility and intermittency on the anomalous scaling. 
Analysis of various simplified models suggests that compressibility strongly affects the passively advected fields. In particular, 
the anomalous exponents become nonuniversal due to dependence on the compressibility parameters, such that the anomalous scaling is enhanced. 

From the RG viewpoints, the key problem is the renormalizability (in plain words, the self-consistency) 
of the model in question. The straightforward RG approach by \cite{6}
suffered by the lack of renormalizability. 
Another approach, proposed in \cite{ANU}, 
with the price of certain reasonable approximations and natural extensions of the original
set-up, lead to an internally consistent renormalizable model in terms of an effective density field.
The model \cite{ANU} exhibits an unique IR fixed point of the RG equation, which allows for detailed one-loop analysis of the IR anomalous scaling behaviour within the RG+SDE approach for the scalar \cite{AK} and magnetic \cite{AK1} fields.\footnote{The crossover between the turbulent advection and ``shaking'' requires special analysis in the vicinity of $d=4$; see \cite{AK2}-\cite{AK5}.
Another special value is $d=2$, where the NS turbulence becomes close to the equilibrium distribution.
The vicinity of $d=2$ requires additional analysis even for the incompressible case; see, e.g. \cite{RedBook}. In the following, we assume $2<d<4$.}

In this paper, we study the model of a transverse vector field (e.g. magnetic)  with the most general form of the nonlinearity, known as the ${\cal A}$ model, 
passively advected by a strongly compressible turbulent flow. The latter is governed by the randomly stirred NS equation and treated according to the approach proposed \cite{ANU} and developed in \cite{AK,AK1}.
Thus, the special case ${\cal A}=1$ corresponds to the kinematic magnetic model \cite{AK1}.

The full stochastic problem is equivalent to a renormalizable field theoretic model with an infrared (IR) attractive fixed point.
Thus, the scaling behaviour for the large-scale, long-distance behaviour is established.
However, the question whether the parameter ${\cal A}$ tends to a certain 
fixed-point value or remains free, cannot be answered within our one-loop approximation.

We shall not describe the contents of this paper here,
because it is clear from the following text. 

\section{Description of the model}

Following \cite{ANU,AK,AK1}, we describe the stochastic dynamics of a
compressible fluid by the set of the equations:
\begin{eqnarray}
\nabla_{t} v_{i} &=&
\nu_{0} [\delta_{ik}\partial^{2}-\partial_{i}\partial_{k}]
v_{k}\! +\! \mu_0 \partial_{i}\partial_{k} v_{k} -\!
\partial_{i} \phi\! +\! f_{i}
\label{ANU} \\
\nabla_{t} \phi &=& -c_{0}^{2} \partial_{i}v_{i},
\label{ANU1}
\end{eqnarray}
which are derived from the momentum balance equation and the continuity
equation \cite{Landau}
with the two assumptions: the kinematic viscosity coefficients
$\nu_{0}$ and $\mu_{0}$ are assumed to be constant, that is, independent of
$x=\{t, {\bf x}\}$,
and the equation of state is taken in the simplest form of the linear
relation $(p-\bar p) = c^{2}_{0} (\rho-\bar\rho)$
between the deviations of the pressure $p(x)$ and the
density $\rho(x)$ from their mean values; then the constant $c_{0}$
has the meaning of the (adiabatic) speed of sound.
Here and below, we use the subscript ``0'' to denote the bare (unrenormalized) parameters to distinguish them from their renormalized (ultraviolet finite) analogs.

As the basic (primary) fields we use the velocity field
$\boldsymbol{v} = \{v_i (x)\}$ and,
instead of the density, the scalar field defined as
$\phi(x) = c_{0}^{2} \ln (\rho(x)/\bar \rho)$. Furthermore,
\begin{eqnarray}
\nabla_{t} = \partial_{t} + v_{k} \partial_{k}
\label{Nabla}
\end{eqnarray}
is the Lagrangian (Galilean covariant) derivative, $\partial_{t} = \partial /\partial t$, $\partial_{i} = \partial /\partial x_{i}$, and
$\partial^{2} =\partial_{i}\partial_{i}$ is the Laplace operator.
The problem is studied in the
$d$-dimensional (for generality) space ${\bf x}=\{x_i\}$, $i=1\dots d$,
and the summations over the repeated Latin indices are always implied.

In the NS equation (\ref{ANU}), $f_{i}$ is the density of the
external force, which models the energy input into the
system from the large-scale stirring. In order to apply the standard
 RG to the problem, and to ensure the Galilean symmetry, the force is taken to be 
 Gaussian with zero mean,
not correlated in time, with the given covariance
\begin{eqnarray}
\langle f_{i}(x) f_{j}(x') \rangle = \delta(t-t') \int_{k>m} \frac{d{\bf k}}
{(2\pi)^{d}} \, D^{f}_{ij}({\bf k}) \exp\{{\rm i} {\bf k}\cdot{\bf x}\},
\nonumber \\ {}
\label{force}
\end{eqnarray}
where
\begin{eqnarray}
D^{f}_{ij}({\bf k}) = D_{0}\, k^{4-d-y}\,
\left\{ P^{\bot}_{ij} ({\bf k})
+ \alpha P^{\parallel}_{ij} ({\bf k}) \right\}.
\label{power}
\end{eqnarray}
Here $P^{\bot}_{ij} ({\bf k})=\delta_{ij}-k_{i}k_{j}/k^{2}$ and
$P^{\parallel}_{ij} ({\bf k})=k_{i}k_{j}/k^{2}$ are the transverse and the
longitudinal projectors, respectively, $k=|{\bf k}|$ is the wave number (momentum), $D_{0}$ and $\alpha$ are positive amplitudes.
The parameter $g_{0}=D_{0}/\nu_0^{3}$ plays the part of the coupling constant (expansion parameter in the perturbation theory); the relation $g_{0} \sim \Lambda^{y}$ defines the typical ultraviolet (UV) momentum scale. The parameter $m \sim L^{-1}$, reciprocal of the integral turbulence scale,
provides IR regularization; its precise form is unessential and the sharp cut-off is merely the simplest choice for the calculation reasons.

The exponent $0<y\le 4$ plays the role analogous to $\varepsilon=4-d$
in the RG theory of critical state \cite{Book}: it provides UV
regularization (so that the UV divergences have the form of the poles in
$y$) and various scaling dimensions are calculated as series in $y$.
The most realistic (physical) value is given by the limit $y\to4$: then
the function (\ref{power}) can be viewed as a power-like representation
of the function $\delta({\bf k})$ and it corresponds to the idealized
picture of the energy input from infinitely large scales.

As already mentioned, more detailed justification and
discussion of the compressible model (\ref{ANU})--(\ref{power})
is given in \cite{ANU,AK,AK1}.

In this paper, we confine ourselves with the case of a transverse (divergence-free) passive vector
$\theta_{i} (x)$ field.
Then the general advection-diffusion equation acquires the form
\begin{eqnarray}
\nabla _t\theta_{i} - {\cal A}_{0} (\theta_{k}\partial_{k}) v_{i} +
\partial_{i} {\cal P} = \kappa_0\partial^{2} \theta_{i} + \eta_{i},
\qquad
\nabla_t \equiv \partial _t + (v_{k}\partial_{k}) ,
\label{1}
\end{eqnarray}
where $\nabla_t$ is the covariant derivative \ref{Nabla},
${\cal P}(x)$ is the 
analog of the pressure, $\kappa_0$ is the diffusivity,
$\partial^{2}$ is the Laplace operator and $\eta_{i}(x)$ is a transverse
Gaussian stirring force with zero mean and covariance
\begin{equation}
\langle \eta_{i}(x) \eta_{k}(x')\rangle = \delta(t-t')\, C_{ik}(r/L).
\label{2}
\end{equation}
The parameter $L$ is an integral scale related to the stirring, and $C_{ik}$
is a dimensionless function with the condition $\partial_{i}C_{ik}=0$, finite
at $r=0$ and rapidly decaying for $r\to\infty$; its precise form is
unessential.

From the physics viewpoints most interesting is the special case
${\cal A}_{0}=1$, where the pressure term disappears: it corresponds to
magnetohydrodynamic (MHD) turbulence. It was studied earlier
within the context of anomalous scaling in
numerous papers; see e.g. \cite{Verg,Rog}, \cite{LM}-\cite{5}
and references therein.

According to the general theorem (see e.g. \cite{Book}), the full-scale
stochastic problem (\ref{ANU}), (\ref{ANU1}), (\ref{force}), (\ref{1}), (\ref{2}),    is equivalent
to the field theoretic model of the doubled set of fields
$\Phi=\{v,v',\phi', \phi',\theta,\theta'\}$ with the action functional
\begin{equation}
{\cal S} (\Phi )= {\cal S}_{v}({ v}', { v}) +
\theta' D_{\theta} \theta'/2 +
\theta' \left\{ -\nabla_{t} - {\cal A}_{0} (\theta_{k}\partial_{k}) v_{i}
+ \kappa_{0} \partial^{2} \right\} \theta,
\label{action}
\end{equation}
where
\begin{eqnarray}
{\cal S}_{v}({ v}', { v}, \phi', \phi) &=& \frac{1}{2} v_{i}'D^{f}_{ik}  v_{k}' +
v_{i}' \left\{ -\nabla_{t} v_{i} +
\nu_{0} \left[\delta_{ik}\partial^{2}-\partial_{i}\partial_{k}\right] v_{k} -\partial_i \phi
\right\}+
\\ \nonumber
&+& \phi' \left[-\nabla_{t}\phi + v_0\nu_0\partial^2 \phi -c_0^2 (\partial_i v_i)\right].
\label{Sv}
\end{eqnarray}

The field theoretic formulation means that various correlation functions
and response (Green) functions of the original stochastic problem are
represented by functional averages over the full set of fields with weight
$\exp {\cal S}(\Phi)$, and in this sense they can be viewed as the Green
functions of the field theoretic model with action (\ref{action}). The
model 
\ref{Sv}
corresponds to standard Feynman diagrammatic techniques with two
vertices $-v'(v\partial)v$ and $-\phi'(v\partial)\phi$ and the free (bare)
propagators, determined by the quadratic part of the action functional; in
the frequency--momentum ($\omega$--${\bf k}$) representation, they have the
forms:
\begin{eqnarray}
\label{propagators}
\langle vv' \rangle_{0} &=& \langle v'v \rangle_{0}^{*}= P^{\bot}
\epsilon_{1}^{-1} + P^{\parallel} \epsilon_{3} R^{-1} ,
\nonumber \\
\langle vv \rangle_{0} &=&  P^{\bot} \frac{d^{f}}{|\epsilon_{1}|^{2}}
+ P^{\parallel}\alpha d^{f} \left|\frac{\epsilon_{3}}{R}\right|^{2},
\nonumber \\
\langle \phi v' \rangle_{0} &=& \langle v' \phi \rangle_{0}^{*}= -
\frac{{\rm i}c_{0}^{2}{\bf k}}{R}, \quad
\langle  v\phi' \rangle_{0} = \langle \phi'v \rangle_{0}^{*}=
\frac{{\rm i}{\bf k}}{R},
\nonumber \\
\langle  \phi\phi' \rangle_{0} &=& \langle \phi'\phi \rangle_{0}^{*}=
\frac{\epsilon_{2}}{R}, \quad
\langle  \phi\phi \rangle_{0} = \frac {\alpha c_{0}^{4} k^{2}d^{f}}
{|R|^{2}},
\nonumber \\
\langle  v\phi \rangle_{0} &=& \langle  \phi v \rangle_{0}^{*} =
\frac{ {\rm i}\alpha c_{0}^{2} d^{f}\epsilon_{3} {\bf k}} {|R|^{2}} ,
\nonumber \\
\langle  \phi'\phi' \rangle_{0} &=& \langle  v'\phi' \rangle_{0} =
\langle  v'v' \rangle_{0} = 0, 
\end{eqnarray}
where we have omitted the vector indices of the fields and the projectors and denoted
\begin{eqnarray}
\epsilon_{1}  &=& -{\rm i}\omega +\nu_0 k^{2}, \quad
\epsilon_{2}=-{\rm i}\omega+ u_0\nu_0 k^{2} ,
\nonumber \\
\epsilon_{3}  &=& -{\rm i}\omega+ v_0\nu_0 k^{2} , \quad
R=\epsilon_{2}\epsilon_{3}+c_{0}^{2}k^{2} ,
\nonumber \\
d^{f} &=& g_{0}\nu_0^{3}\, k^{4-d-y}.
\label{energies}
\end{eqnarray}
The full  model involves additional relevant propagator
\begin{eqnarray}
\label{propagators2}
\langle \theta'_{i}\theta_{j} \rangle_{0}   &=&
\frac{P^{\bot}_{ij} ({\bf k})}
{-{\rm i}\omega+ w_{0} \nu_{0} k^{2} }
\end{eqnarray}
and the new vertex $V_{ijl}\theta'_{i}\theta_{j}v_{l}$ with vertex factor
\begin{equation}
V_{ijl}({\bf k}) = {\rm i} (\delta_{ij}k_{l}- {\cal A}_{0} \delta_{il}k_{j}).
\label{vertex1}
\end{equation}
In the limit $u_{0}\to\infty$, the propagators $\langle vv' \rangle_{0}$
and $\langle vv \rangle_{0}$ become purely transverse, and we arrive at the case of incompressible fluid. Furthermore, the countertems are polynomial in $c_0$ and can be calculated with $c_0=0$ in $R$, which essentially simplifies the calculations; for more details, see \cite{ANU,AK,AK1}.

\section{Renormalization and counterterms}

The analysis based on dimensionality considerations and symmetries shows that the
full extended model (\ref{action}) appears to be multiplicatively renormalizable;
see \cite{ANU,AK,AK1}. 
This means that all the UV divergences can be removed from the Green
functions by the renormalization of the fields $\phi\to Z_{\phi}\phi$,
$\phi'\to Z_{\phi'}\phi'$ and of the parameters:
\begin{equation}
g_{0} = g\mu^y Z_{g}, \quad \nu_0 Z_{\nu}, \quad c_{0} = c Z_{c}\, ,
\label{Ren}
\end{equation}
and so on. Here the renormalization constants $Z_{i}$ absorb all the UV
divergences, so that the Green functions are UV finite (that is, finite at
$y=0$) when expressed in terms of the renormalized parameters $g,u$, and so
on; the reference scale (or the ``renormalization mass'') $\mu$ is an
additional free parameter of the renormalized theory. No renormalization
of the fields $\boldsymbol{v}',\boldsymbol{v}$
and of the parameters $m,\alpha$ is needed.

The renormalized analog of the action functional (\ref{action}) has the form
\begin{equation}
{\cal S}^{R}(\Phi)={\cal S}_{v}^{R}(\Phi) +
{\cal S}_{\theta}^{R}(\Phi),
\label{FactR}
\end{equation}
where
${\cal S}^{R}(\Phi)$ is the renormalized analog of the action
${\cal S}(\Phi)$, given in \cite{ANU,AK,AK1},
\begin{eqnarray}
{\cal S}_{\theta}^{R} &=& \theta'_{i} \bigl\{
-\partial _t\theta_{i} - \partial_{k} (v_{k}\theta _{i} - Z_A {\cal A}\theta_{k} v_{i})
+ \kappa Z_{\kappa} \partial^{2} \theta_{i}
\bigr\}, \nonumber \\
{\cal S}^{R}(\Phi) &=& \frac{1}{2} v_{i}'D^{f}_{ik}  v_{k}' +
v_{i}' \left\{ -\nabla_{t} v_{i} +
Z_{1} \nu [\delta_{ik}\partial^{2}-\partial_{i}\partial_{k}] v_{k} +
Z_{2} u \nu\partial_{i}\partial_{k} v_{k} - Z_{4}\partial_{i} \phi \right\} +
\nonumber \\
&+& \phi' \left[ -\nabla_{t}\phi  + Z_{3}v \nu \partial^{2} \phi -
Z_{5} c^{2}(\partial_{i}v_{i}) \right].
\label{Rak}
\end{eqnarray}

 The renormalized action (\ref{Rak}) is obtained from the original one (\ref{action})
by the renormalization of the fields $\phi\to Z_{\phi}\phi$,
$\phi'\to Z_{\phi'}\phi'$ and the parameters
\begin{eqnarray}
g_{0} &=& g \mu^y Z_{g}, \quad \nu_0=\nu Z_{\nu}, \quad
u_{0}= u Z_{u},
\nonumber \\
v_{0} &=& v Z_{v}, \quad c_{0}= c Z_{c},
\nonumber \\
{\cal A}_0 &=& Z_A {\cal A}, \quad \kappa_0 = Z_{\kappa} \kappa.
\label{mult}
\end{eqnarray}
The renormalization constants in (\ref{Rak}) and (\ref{mult}) are
related as
\begin{eqnarray}
Z_{\nu} &=& Z_{1}, \quad Z_{u}=Z_{2}Z_{1}^{-1},
\nonumber \\
Z_{v} &=&  Z_{3}Z_{1}^{-1}, \quad Z_{\phi}=Z_{\phi'}^{-1}=Z_{4},
\nonumber \\
 Z_{c} &=& (Z_{4}Z_{5})^{1/2}, \quad Z_{g}= Z_{\nu}^{-3}.
\label{relat}
\end{eqnarray}

In this paper we present results of the one-loop calculation of the renormalization constants $Z_{\kappa}$ and $Z_{\cal A}$.

\section{One-loop calculation and Galilean symmetry}

\subsection{Triangle diagrams}

In order to calculate the renormalization constant $Z_{\cal A}$ we employed the MS scheme to eliminate the divergent parts for all the Feynman diagrams which correspond to the Green function $\langle \theta'\theta v \rangle$. In this manuscript we confined to the first order (one-loop approximation).
For the readers who are interested in the calculations details we present the 
one-loop answers for UV divergent parts 
(that are, poles in $y$) for the three diagrams:
\begin{eqnarray}
\raisebox{-5.ex}{\includegraphics [width=.1\textwidth,clip]{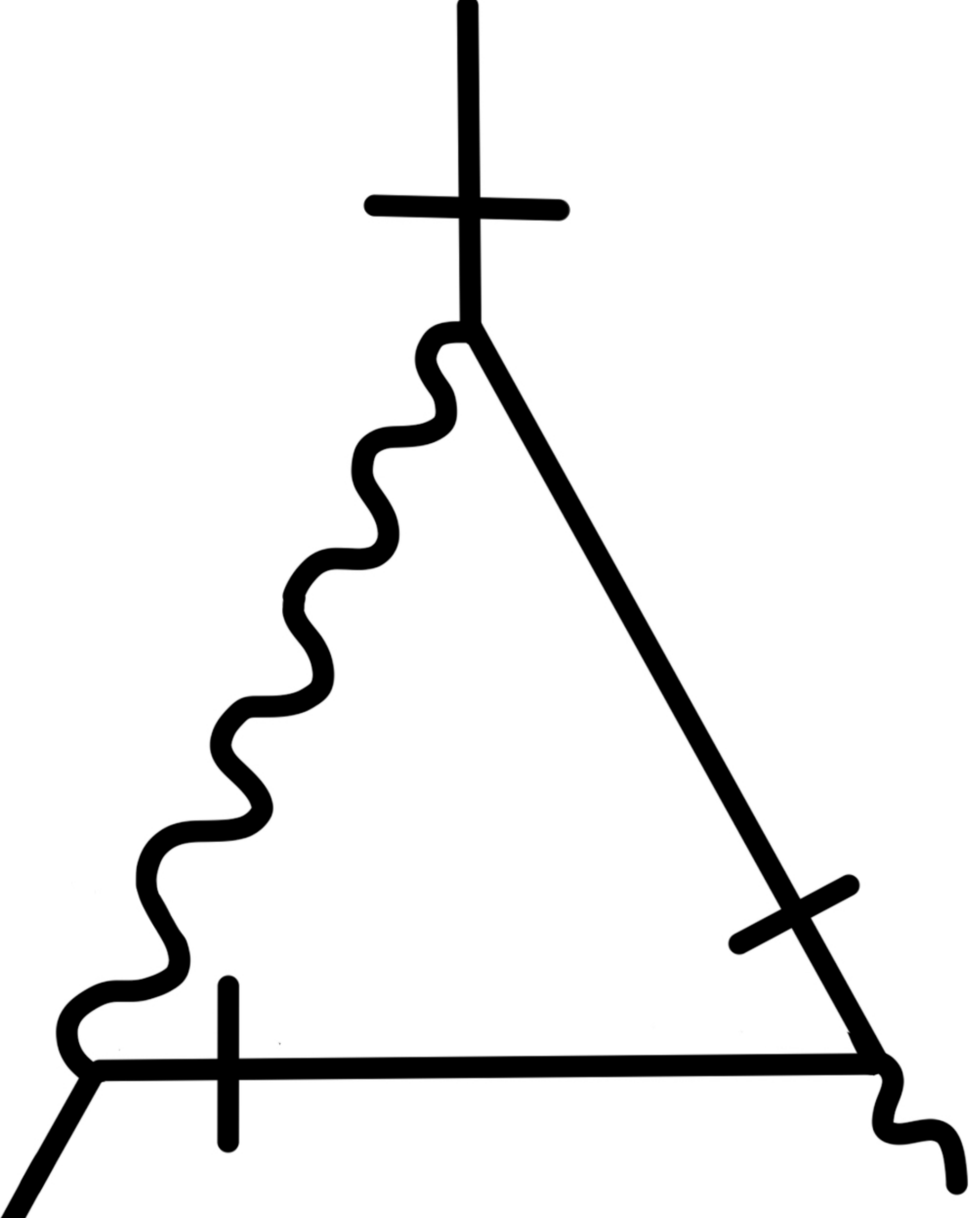}} &=& \frac{i \alpha}{2ud(u+w)^2}(p_3\delta_{12} - {\cal A} p_2\delta_{13}) + 
i ({\cal A}-1) \frac{p_2 \delta_{13} + p_3 \delta_{12}}{d(d+2)} \Big( \frac{-{\cal A}}{2\nu^3(w+1)^2} + \frac{\alpha}{2u(u+w)^2} \Big),
\label{V1} \\
\raisebox{-5.ex}{\includegraphics [width=.1\textwidth,clip]{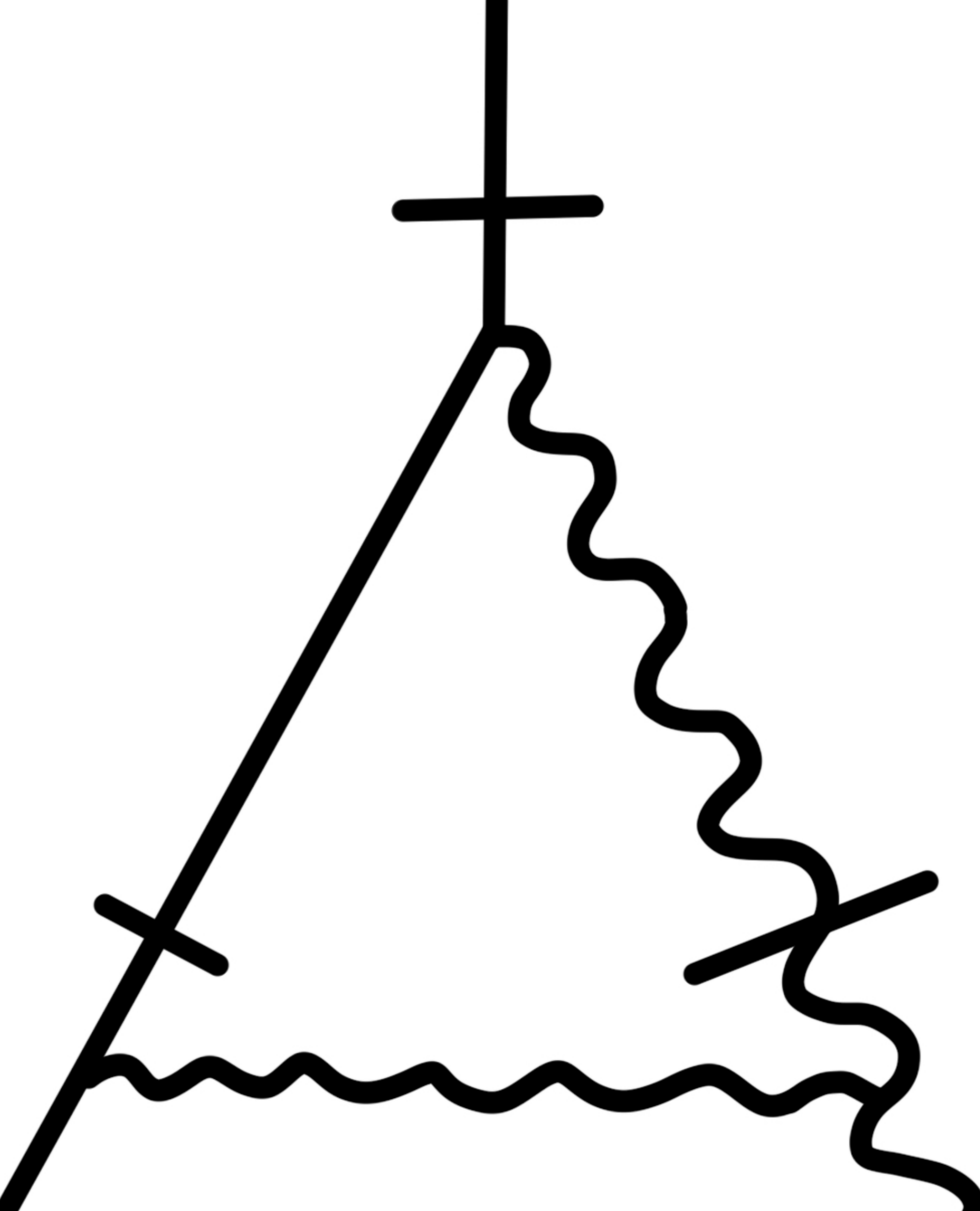}} &=& 
\frac{-i{\cal A}(1-{\cal A})(w+3)}{4\nu^3(w+1)^2} 
\frac{p_2 \delta_{13} + p_3 \delta_{12}}{d(d+2)}
+
\frac{-i\alpha(3u+w)}{4u^2(w+u)^2}
\Big[
p_3\delta_{12}\frac{{\cal A}+d+1}{d(d+2)} - 
p_2\delta_{13}\frac{{\cal A}d+{\cal A}+1}{d(d+2)}
\Big],
\label{V2} \\
\raisebox{-5.ex}{\includegraphics [width=.1\textwidth,clip]{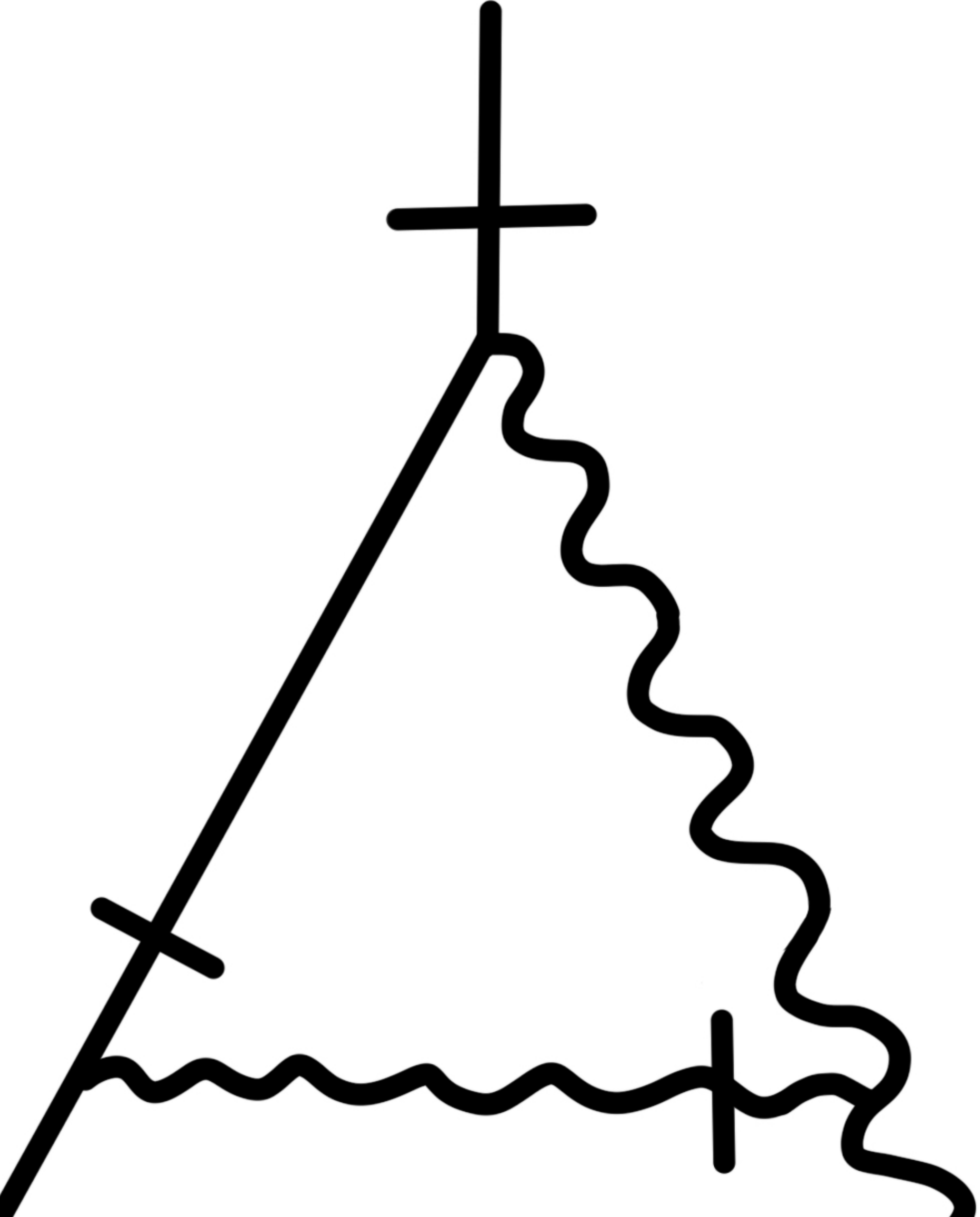}} &=& 
\frac{i\alpha}{4u^2\nu^3(u+w)}
\Big[
\frac{-p_2 \delta_{13} + (d+1) p_3 \delta_{12}}{d(d+2)} - \frac{(d+1)p_2 \delta_{13} - p_3 \delta_{12}}{d(d+2)}
\Big]
+ \frac{i{\cal A}({\cal A}-1)}{4\nu^3(w+1)}\frac{p_2 \delta_{13} + p_3 \delta_{12}}{d(d+2)},
\label{V3}
\end{eqnarray}
with the common overall factor $g\mu^y$.

It is easy to verify that the sum of the above 
expressions vanishes, that is, the sum of the three one-loop diagrams is UV finite.

In the rapid-change version of our model, $Z_{\cal A}=1$ and
$\gamma_{\cal A}=0$ identically because all nontrivial Feynman diagrams of the 1-irreducible Green function $\langle\theta'v\theta\rangle$ contain closed circuits of retarded propagators and therefore vanish; see, e.g., the discussion in \cite{2}.

In more realistic cases, the absence of the $O(g)$ term in $Z_{\cal A}$ and $\gamma_{\cal A}$
results from the cancellation of the (nontrivial) contributions from
the three one-loop diagrams in the 1-irreducible Green function
$\langle\theta'v\theta\rangle$. For the counterterm
$\theta' (v\partial) \theta$ such a cancellation is guaranteed by the
Galilean symmetry (to all orders in $g$). 

For the magnetic case ${\cal A}=1$ we have $Z_1 = Z_2=1$ 
and, therefore, $Z_ {\cal A}= 1$, to all orders in $g$.
There, this is a consequence of the transversality relation
$\partial_i V_i=0$ for the vertex
$V_i = \partial_k (v_i \theta_k - v_k \theta_i)$.

For general ${\cal A}\ne1$ and incompressible fluid,  the cancellation of the terms $\theta'(\theta\partial)v$ looks accidental and could be explained {\it a posteriori} 
by a rather simple form of the one-loop diagrams: the structures corresponding to the counterterm $\theta' (v\partial) \theta$ cancel each other due to the Galilean symmetry, 
while the structures corresponding to $\theta'(\theta\partial)v$ enter all the one-loop
diagrams with the same coefficients and cancel out into the bargain \cite{4}. However, 
this explanation can hardly pass beyond the one-loop approximation even for the 
incompressible fluid; thus nontrivial contributions of the order $g^{2}$
and higher in $Z_{\cal A}$ and $\gamma_{\cal A}$ are not forbidden.
Furthermore, in our compressible case, the expressions for the one-loop triangle diagrams become much more complicated, and their accidental
 cancellation raises serious doubts. However, a possible deep reason behind this effect remains uncovered.

\subsection{Self-energy $\Sigma$} 

The constant $Z_{\kappa}$ is found from the requirement for the 1-irreducible Green
function $\langle\theta'\theta\rangle_{\rm 1-ir}$ be UV finite (that is,
finite at $y\to0$) when expressed in renormalized parameters. In the
frequency--momentum representation it has the form:
\begin{equation}
\langle\theta'_{1}\theta_{2}\rangle_{\rm 1-ir}(\Omega,{\bf p}) =
\left\{- \kappa_0 p^{2} + {\rm i}\Omega  \right\}
\, P_{12}^{\bot}({\bf p})
+ \Sigma_{12} (\Omega,  {\bf p}),
\label{Dyson}
\end{equation}
where  $\Sigma_{12}$ is the ``self-energy operator'' given by infinite sum
of 1-irreducible Feynman diagrams and $p=|{\bf p}|$.
Because of the large number of tensor
indices involved in our expressions, we use numbers (instead of latin
letters) to denote them, with the standard convention on the summation over
repeated indices.

The only one-loop self-energy diagram looks as follows:
\begin{equation}
\Sigma_{12} = \raisebox{-2.ex}{\includegraphics [width=.1\textwidth,clip]{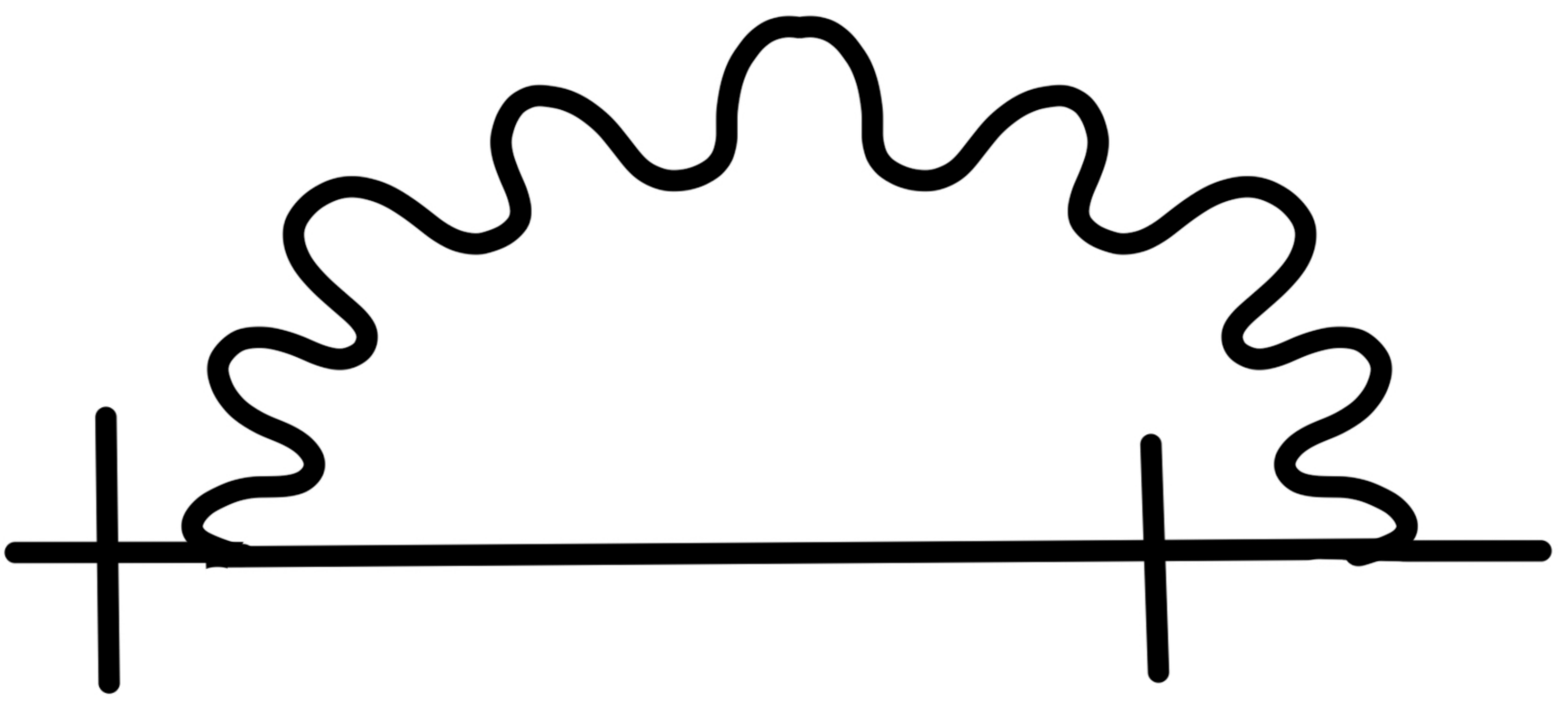}},
\label{Self}
\end{equation}
 were the wavy line denotes the propagator $\langle vv \rangle$, the straight line crossed out at the right end denotes the propagator $\langle \theta \theta' \rangle$, and the vorticies composed of one wavy tip and two straight tips with one crossing correspond to the Green function $V_{123} =\langle v_1 \theta_2 \theta'_3 \rangle $.

Here we do not present the details of the calculations
and give only the result:
\begin{eqnarray}
\Sigma_{12} &=& - \nu p^2 P^{}_{12}({\bf p}) \frac{{\hat g}}{2dy} \Big( \frac{\mu}{m} \Big )^y  \
 {\bf\Big[}  \frac{1}{1+w}\Big\{ 
d-1 + ({\cal A} - 1)\Big[ \frac{2w{\cal A}}{(1+w)(d+2)} + \frac{{\cal A} + d + 1}{d+2} \Big] \Big\} \nonumber \\
&+& \frac{\alpha}{u + w} \Big\{ \frac{u-w}{u+w} + ({\cal A} - 1)\Big[ \frac{-2w{\cal A}}{(u+w)(d+2)} - \frac{{\cal A} + {\cal A}d + 1}{d+2}\Big]
\Big\} {\bf\Big]}.
\end{eqnarray}
Here we passed to the new coupling constant
\begin{eqnarray}
{\hat g}=g S_{d}/(2\pi)^{d}, \qquad S_{d}=2\pi^{d/2}/\Gamma(d/2)
\label{ghat}
\end{eqnarray}
where $S_{d}$ is the surface area of the unit sphere in $d$-dimensional space.

Then in the MS scheme the renormalization constant $Z_{\kappa}$ that
cancels the pole in the renormalized
analog of the function  (\ref{Dyson}) (that is, with the replacement
$\kappa_{0}\to \kappa Z_{\kappa}$ in the bare term) has the form:
\begin{eqnarray}
Z_{\kappa} = 1 &-&  \frac{g}{2dyw}{\bf\Big[}  \frac{1}{1+w}\Big\{ 
d-1 + ({\cal A} - 1)\Big[ \frac{2w{\cal A}}{(1+w)(d+2)} + \frac{{\cal A} + d + 1}{d+2} \Big] \Big\} \nonumber \\
&+& \frac{\alpha}{u + w} \Big\{ \frac{u-w}{u+w} + ({\cal A} - 1)\Big[ \frac{-2w{\cal A}}{(u+w)(d+2)} - \frac{{\cal A} + {\cal A}d + 1}{d+2}\Big]
\Big\} {\bf\Big]},
\end{eqnarray}
while the corresponding anomalous dimension is
\begin{eqnarray}
\gamma_{\kappa} &=& \frac{g}{2dw}{\bf\Big[}  \frac{1}{1+w}\Big\{ 
d-1 + ({\cal A} - 1)\Big[ \frac{2w{\cal A}}{(1+w)(d+2)} + \frac{{\cal A} + d + 1}{d+2} \Big] \Big\} \nonumber \\
&+& \frac{\alpha}{u + w} \Big\{ \frac{u-w}{u+w} + ({\cal A} - 1)\Big[ \frac{-2w{\cal A}}{(u+w)(d+2)} - \frac{{\cal A} + {\cal A}d + 1}{d+2}\Big]
\Big\} {\bf\Big]}
\end{eqnarray}
with the corrections of the order ${\hat g}^{2}$ and higher.

\section{RG equations, $\beta$ functions and fixed points}

Multiplicative renormalizability of the field theoretic model
allows one to derive, in a standard way, differential RG equations for the
renormalized Green functions
\[ G(e,\mu,\dots) = \langle \Phi \dots \Phi \rangle_{R}. \]
Here
$e=\{g,\nu,u,v,w,c,m,\alpha\}$ is the full set of renormalized parameters,
$\mu$ is the reference momentum scale and the ellipsis stands for the other
arguments (times or frequencies and coordinates or momenta). For convenience,
we introduced here three dimensionless ratios: $u_{0}=\mu_0/\nu_0$ and
$v_{0}=\chi_0/\nu_0$ are related to the viscosity and diffusivity coefficients
of the (properly extended) model (\ref{ANU}), (\ref{ANU1}), while
$w_{0}=\kappa_0/\nu_0$ is related to the magnetic diffusivity coefficient;
$u,v,w$ are their renormalized analogs.

The RG equation expresses the invariance of the renormalized Green function
with respect to changing of the reference scale $\mu$, when the bare
parameters $e_{0}$ are kept fixed:
\begin{equation}
\left\{ \widetilde{\cal D}_{\mu} + \sum_{\Phi} N_{\Phi}\gamma_{\Phi}
\right\} \,G(e,\mu,\dots) = 0.
\label{RGE}
\end{equation}
Here and below we denote ${\cal D}_{x} \equiv x\partial_{x}$ for any
variable $x$ and $\widetilde{\cal D}_{\mu}$ is the operation
${\cal D}_{\mu} \equiv \mu\partial_{\mu}$ at fixed $e_{0}$. In terms of
the renormalized variables, it takes the form
\begin{equation}
\widetilde{\cal D}_{\mu}  = {\cal D}_{\mu} + \beta_{g}\partial_{g} +
\beta_{u}\partial_{u} + \beta_{v}\partial_{v}+ \beta_{w}\partial_{w}
- \gamma_{\nu}{\cal D}_{\nu}- \gamma_{c}{\cal D}_{c} .
\label{RG2}
\end{equation}
The anomalous dimension $\gamma_{F}$ of a certain quantity $F$
(a field or a parameter) is defined by the relation
\begin{equation}
\gamma_{F}= \widetilde{\cal D}_{\mu} \ln Z_F ,
\label{RGF1}
\end{equation}
and the $\beta$ functions for the dimensionless parameters (``coupling
constants'') are
\begin{eqnarray}
\beta_{g} &=& \widetilde{\cal D}_{\mu} g = g\,[-y-\gamma_{g}],
\nonumber \\
\beta_{u} &=& \widetilde{\cal D}_{\mu} u = -u\gamma_{u},
\label{betagw}
\end{eqnarray}
and similarly for $\beta_{v}$, $\beta_{w}$. Here the second equalities
result from the definitions and the relations of the type (\ref{Ren}).

Note that from the definition of $w_{0}$ it follows that
$Z_{\kappa}=Z_{\nu}Z_{w}$, so that $\beta_{w} =  w [\gamma_{\nu} - \gamma_{w}]$, and
\begin{eqnarray}
\label{betaw}
    \beta_w = w y \Big[
    \frac{1}{3} - \frac{1}{w(w+1)} 
    \Big(
    \frac{2}{3} + \frac{\alpha (1-w)}{3(1+w)} + ({\cal A}-1)(1-\alpha) \frac{2w{\cal A} + (w+1)({\cal A}+4)}{15(w+1)} - \frac{\alpha}{5}({\cal A}-1)^2
    \Big)
    \Big].
\end{eqnarray}
For better transparency, here and  below we put $d=3$, unless stated otherwise.

The possible types of IR asymptotic behavior are associated with IR
attractive fixed points of the RG equations. The coordinates
$g_{*}=\{g_{i*}\}$ of the fixed points are found from the equations
\begin{equation}
\beta_{i} (g_{*}) =0,
\label{points}
\end{equation}
where $g=\{g_i\}$ is the full set of coupling constants and
$\beta_{i}=\widetilde{\cal D}_{\mu}g_i$ are their $\beta$ functions.
The character of a fixed point is determined by the matrix
\begin{equation}
\Omega_{ij}=\partial\beta_{i}/\partial g_{j}|_{g=g_{*}}.
\label{Omega}
\end{equation}
For the IR fixed points the matrix $\Omega$ is positive
(that is, positive are the real parts of all its eigenvalues).

The analysis performed in~\cite{ANU} (see also \cite{AK,AK1}) on the base
of the leading-order (one-loop) approximation has shown that the RG
equations of the model ${\cal S}_{v}$, corresponding to the stochastic NS
problem (\ref{ANU}), (\ref{ANU1}),
possess the only IR attractive fixed point in the physical region of
parameters ($g,u,v>0$)\footnote {It was argued in \cite{8} 
that these first-order fixed-point coordinates \cite{ANU}
do not satisfy certain thermodynamic stability inequalities \cite{Landau}. 
We avoid discussion of this interesting issue here and only refer to the two-loop results \cite{86} which do not contradict the stability.}:
\begin{equation}
\hat g_{*} = \frac{4dy}{3(d-1)} +O(y^{2}) = 2y+O(y^{2}), \quad u_{*}=1+O(y), \quad
v_{*}=1+O(y).
\label{FP}
\end{equation}

From a certain exact relation between the renormalization constants
\cite{ANU}, the exact result
\begin{eqnarray}
\gamma_{\nu}^{*}=y/3
\label{Anom}
\end{eqnarray}
follows (no corrections of the order $y^{2}$ and higher). Here and below,
$\gamma_{i}^{*}$ denotes the value of the anomalous dimension
$\gamma_{i}$ at the fixed point.

Now we substitute the one-loop expressions 
and the exact result (\ref{Anom}) into Eq. (\ref{betaw}). Then the
equation $\beta_{w} =0$ yields the equation
\begin{eqnarray}
         \frac{2}{3} 
                  + \frac{\alpha (1-w)}{3(1+w)} + ({\cal A}-1)(1-\alpha) \frac{2w{\cal A} + (w+1)({\cal A}+4)}{15(w+1)} - \frac{\alpha}{5}({\cal A}-1)^2
     = \frac{w(w+1)}{3}.
    \label{eq}
\end{eqnarray}
To find the coordinates of the fixed points one has to find the roots of 
the cubic in-$w$ equation (\ref{eq}), the problem solved by the Cardano formula.
In general case, the equation (\ref{eq}) has at least one real root,
the other two can be real or complex depending on the model parameters.
Furthermore, we are interested in IR attractive fixed points.
The functions $\beta_{g,u,v}$ do not depend on $w$, so that the new eigenvalue
of the matrix (\ref{Omega}) coincides with the diagonal element
$\partial\beta_{w}/\partial w|_{g=g_{*}}$.

Thus, the full description of the pattern of the fixed points (and their stability regions) in the full space of model parameters appears rather complicated.
Below we will briefly touch only a few most interesting representative
special cases.

\begin{itemize}
    \item ${\cal A} = 1$:  The only real root of the equation (\ref{eq}) is $w_* = 1$. It turns out that the point is IR-attractive ($\partial\beta_{w}/\partial w|_{g=g_{*}}~>~0$). This case corresponds to the magnetic field passively advected by the compressible fluid (kinematic MHD approximation). This model was earlier considered in \cite{AK1}; the are in agreement.
\end{itemize}
       
\begin{itemize}
    \item  ${\cal A} = 0$: The new symmetry appears: the model is invariant with respect to the shift $\theta \to \theta +$+ const. 
    For $\alpha > 0$ there is always an IR attractive fixed point, with the coordinate depending on $\alpha$. Mention an interesting special case within this special case: when $\alpha = 4$ and ${\cal A} = 0$, one obtains 
    $w_{*} = 1$. 
    
    It also turns out, that for some negative values of $\alpha$ (for example, $-1.1< \alpha< -1.05$) there are two real fixed points, but only one of them is IR attractive. Of course, negative values of $\alpha$ require inventive physical interpretation.
    
    \item  ${\cal A} = -1$: For $0~<~\alpha~<~4$ there is always an IR attractive fixed point, whose coordinates depend on $\alpha$. This case 
    can be interpreted as a linearized equation for the perturbation $\theta$ around the background velocity field $v$; see, e.g. discussion in \cite{2}.
   \end{itemize}

Existence of an IR attractive fixed point in the physical region of the
parameters implies existence of scaling behavior in the IR range.
The critical dimension of some quantity $F$ (a field or a parameter)
is given by the relation (see \cite{Book,UFN,RedBook})
\begin{equation}
\Delta_{F} = d^{k}_{F}+ \Delta_{\omega}d^{\omega}_{F} + \gamma_{F}^{*},
\quad
\Delta_{\omega} =  2-\gamma_{\nu}^{*} = 2-y/3.
\label{Krit}
\end{equation}
Here $d^{k}_{F}$ and $d^{\omega}_{F}$ are the canonical dimensions of $F$,
$\gamma_{F}^{*}$ is the value of the anomalous dimension $\gamma_{F}$
at the fixed point,
and $\Delta_{\omega}$ is the critical dimension of the frequency.

The critical dimensions of the fields and parameters of the model described
by the action ${\cal S}_{v}$ from Eq. (\ref{Sv}) are presented
in~\cite{ANU}; see also ~\cite{AK,AK1}:
\begin{equation}
\Delta_{v}=1-y/3, \quad \Delta_{v'}= d- \Delta_{v}, \quad
\Delta_{\omega}=2-y/3, \quad \Delta_{m}=1
\label{KritEx}
\end{equation}
(these results are exact due to $\gamma_{\nu}^{*}=y/3$ and
$\gamma^{*}_{v,v',m}=0$) and
\begin{equation}
\Delta_{\phi}=d-\Delta_{\phi'}=2-5y/6+O(y^{2}), \quad
\Delta_{c}= 1- 5y/12 +O(y^{2}).
\label{Krit2}
\end{equation}
In addition, our full model involves
two more critical dimensions, whose meanings can be found in~\cite{AK2}:
\begin{equation}
\Delta_{\theta}= -1+y/6, \quad \Delta_{\theta'}= d+1 -y/6.
\label{KriTet}
\end{equation}
These expressions are exact  for general $d$ because the fields ${\theta}$ and ${\theta}'$ are not renormalized.

\section{Conclusion and open problems}

We have studied a stochastic model of a transverse (divergence-free, e.g., magnetic) vector field $\theta$, passively advected by a random non-Gaussian velocity field with finite correlation time, governed by the 
stochastic NS equations for a strongly compressible fluid. The model is described by an advection-diffusion equation with a random large-scale stirring force, nonlocal pressure term and the most general form of the inertial nonlinearity, ``controlled'' by the parameter ${\cal A}$. The NS model is treated according to the approach
advocated in \cite{ANU,AK1,AK2}. Within this approach,
an extended full-scale model appears renormalizable, and the RG techniques can be applied to its IR behaviour.

The full model reveals a rather complicated manifold of the possible types of asymptotic behaviour and crossover regimes, governed by various fixed points 
of the RG equations.
Existence, among them, of IR attractive points shows that 
the correlation functions of the model fields may exhibit scaling behaviour
for certain areas of the parameters ${d,y,\alpha,{\cal A}}$. 

The corresponding scaling dimensions for the basic fields and parameters are
given by the one-loop order exactly (that is, to all orders in the $y$ expansion).

The open question is whether the amplitude $\cal{A}$ in front of the ``stretching term'' $(\theta\partial)v$ 
in the advection-diffusion equation tends to some fixed-point value, 
or it survives as a free parameter which the anomalous  dimensions depend upon. 
The solution remains elusive: in the one-loop approximation, the practical calculation shows that the 
fixed-point values appear arbitrary,
as the consequence of the relation $Z_{A}=1$.
At the same time, any kind of
identities, expressing the Galilean symmetry, impose no 
restriction on the exact fixed-point value of $Z_{A}$.
Hopefully, the practical two-loop calculation could resolve 
this dilemma. 

As the next steps, the two-loop renormalization of the models in question should be undertaken.
Another following problem is the 
inclusion of the relevant composite operators, 
responsible for  the anomalous multiscaling, 
The work is already in progress.

\section*{Acknowledgments} 

The authors are indebted to L.Ts.~Adzhemyan, N.M.~Gulitskiy, M.~Hnatich, J.~Honkonen and P.I.~Kakin for valuable discussions.

The reported study was funded by the Russian Foundation for Basic Research~(RFBR), project number~19-32-60065. 
Maria M. Tumakova was also supported by the Foundation for the Advancement of Theoretical Physics and Mathematics ``BASIS''.


\end{document}